\documentclass[aps,pre,floatfix,10pt]{revtex4}
\usepackage[utf8]{inputenc}
\usepackage{amsmath}
\usepackage{graphicx}
\usepackage{float}
\usepackage{xcolor}
\usepackage{amssymb}
\usepackage{bm}

\begin{document}

\title{{\bf Decentralized decision making by an ant colony:\\ drift-diffusion model of individual choice, quorum and collective decision {\footnote{The first two authors contributed equally.} }}}
\author{Smriti Pradhan} 
\affiliation{Department of Physics, Indian
  Institute of Technology Kanpur, 208016, India} 
\author{Swayamshree Patra} 
\affiliation{Department of Physics, Indian
  Institute of Technology Kanpur, 208016, India} 
\author{Debashish Chowdhury{\footnote{E-mail: debch@iitk.ac.in (Corresponding author)}}}
\affiliation{Department of Physics, Indian Institute of Technology
  Kanpur, 208016, India}

\begin{abstract}
Ants are social insects. When the existing nest of an ant colony becomes uninhabitable, the hunt for a new suitable location for migration of the colony begins. Normally, multiple sites may be available as the potential new nest site. Distinct sites may be chosen by different scout ants  based on their own assessments. Since the individual assessment is error prone, many ants may choose inferior site(s). But, the collective decision that emerges from the sequential and decentralized decision making process is often far better. We develop a model for this multi-stage decision making process. A stochastic drift-diffusion model (DDM) captures the sequential information accumulation by individual scout ants for arriving at their respective individual choices. The subsequent tandem runs of the scouts, whereby they recruit their active nestmates, is modelled in terms of suitable adaptations of the totally asymmetric simple exclusion processes (TASEP). By a systematic analysis of the model we explore the conditions that determine the speed of the emergence of the collective decision and the quality of that decision.
More specifically, we demonstrate that collective decision of the colony is much less error-prone that the individual decisions of the scout ants. We also compare our theoretical predictions with experimental data.

\end{abstract}

\maketitle

\section{Introduction}

Decision making by an individual organism, including humans, by sequential accumulation of relevant information \cite{evans19}, is a stochastic process. The dynamics of this cognitive process \cite{wardbook} have been modelled in the past few decades in terms of drift-diffusion model (DDM) and its various adaptations for different situations \cite{ratcliff16}. In these models a decison is made when the accumulated information reaches a threshold value. However, the speed of attaining a threshold is not the sole criterion for the efficiency of the decision-making process. Given multiple alternatives to choose from, accuracy of the choice, i.e., chosing the best option, is also highly desirable. 

Thus, speed-accuracy tradeoff (SAT) has been one of the main topics of discussion in this context \cite{chittka09,heitz14}. More generally, studies of SAT in sequential processes with rewards \cite{forstmann16,clithero18} manifests as the ``explore-versus-exploit'' dilemma \cite{addicott17}. 
On the one hand, additional information gathering through further exploration, instead of relying on the exploitation of information already gathered, is expected to increase the accuracy of decision making although it delays decision thereby raising the ``cost'' of the decision making process. On the other hand, exploitation of the available information, stopping further exploration, to arrive at a decision can make the decision making faster, thereby reducing the cost, although the risk of wrong decision would be higher.

It has been known for quite some time that the collective decision of colonies of social insects, like ants, is often much better than the quality of the individual decision of the insects \cite{sasaki18}. Each social insect colony, like that of ants, can be regarded as a `super-organism' \cite{holldoblerbookSupOrg}. Such super-organisms, exhibit `swarm intelligence' \cite{kennedybook,bonabeaubook} and make better decision than that of  a single organism in spite of the fact that the super-organism does not possess any brain-like central information-processing unit. What makes the collective decision-making by an ant colony very interesting from the perspective of statistical physics is the interplay of stochastic cognitive dynamics of the individual ants and their interactions in the emergence of the collective final decision. In this paper we present a theoretical model motivated by this phenomenon in the specific context of nest hunting by an ant colony.

\section{House hunting by ant colony: brief description of the phenomenon}

Not all species of ants follow the same strategy for selection of new nest location. In this paper we consider the ant species that utilize the mechanism of ``tandem running'' to recruit more nestmates in one stage of the house hunting process \cite{franklin14}. It is believed that the tandem running strategy is an efficient mechanisms of nest mate recruitment for relatively small colonies of ants \cite{franklin14}. In relatively large ant colonies recruitment of nestmates is accomplished by indirect communication between the ants based on a chemical, called pheromone,  dropped by ants on their trails while foraging. In contrast, recruitment based on tandem running does not use trail pheromone. The mechanism of tandem runs in explained below. 

In the ant species under consideration here, the population of ants in a nest can be categorized into three types, namely, {\it scout} ants, {\it active} ants and {\it passive} ants, based on their task \cite{mallon01, frank02, pratt02,  dornhaus04,pratt05}  (see Fig.\ref{Stage_DM}a).According to the available statistics for {\it Temnothorax Albipennis}, there are a total of 300 ants in a typical nest, out of which 100  are {\it scout ants}, 70  are {\it active ants} and the rest 130 are {\it passive ants} \cite{pratt05} (see Fig.\ref{Stage_DM}(a)). In decision making, the scout ants are the primary assessors who go out and evaluate various nest choices. They are also recruiters for the active ants. Active ants make their decision about potential new homes based on the feedback provided by the scout ants via ``tandem runs'' \cite{moglich78,richardson07,franklin14}. {Tandem run is the phenomenon where one knowledgeable ant leads another naive ant to a potential food source or new nest site. The follower ant,in this case maintains close contact with the lead ant.  When the combined population of the active and scout ants at a new site reaches a threshold the collective decision is made to accept that site as the new nest site for the colony; this phenomenon is often referred to as ``quorum sensing'' which is exhibited by several other unicellular as well as multicellular organisms under wide varieties of circumstances. Once this collective decision is made, the passive ants are physically carried by the scout ants to their new home \cite{pratt02}. The multiple stages of decision making in ant colonies is summarised in Fig.\ref{Stage_DM}(b-d). Thus, the final collective decision about the selection of the new nest emerges from a decentralised process.  In the next section, we frame our model for this collective decision making by an ant colony.

\begin{figure*}
\includegraphics[width=\textwidth]{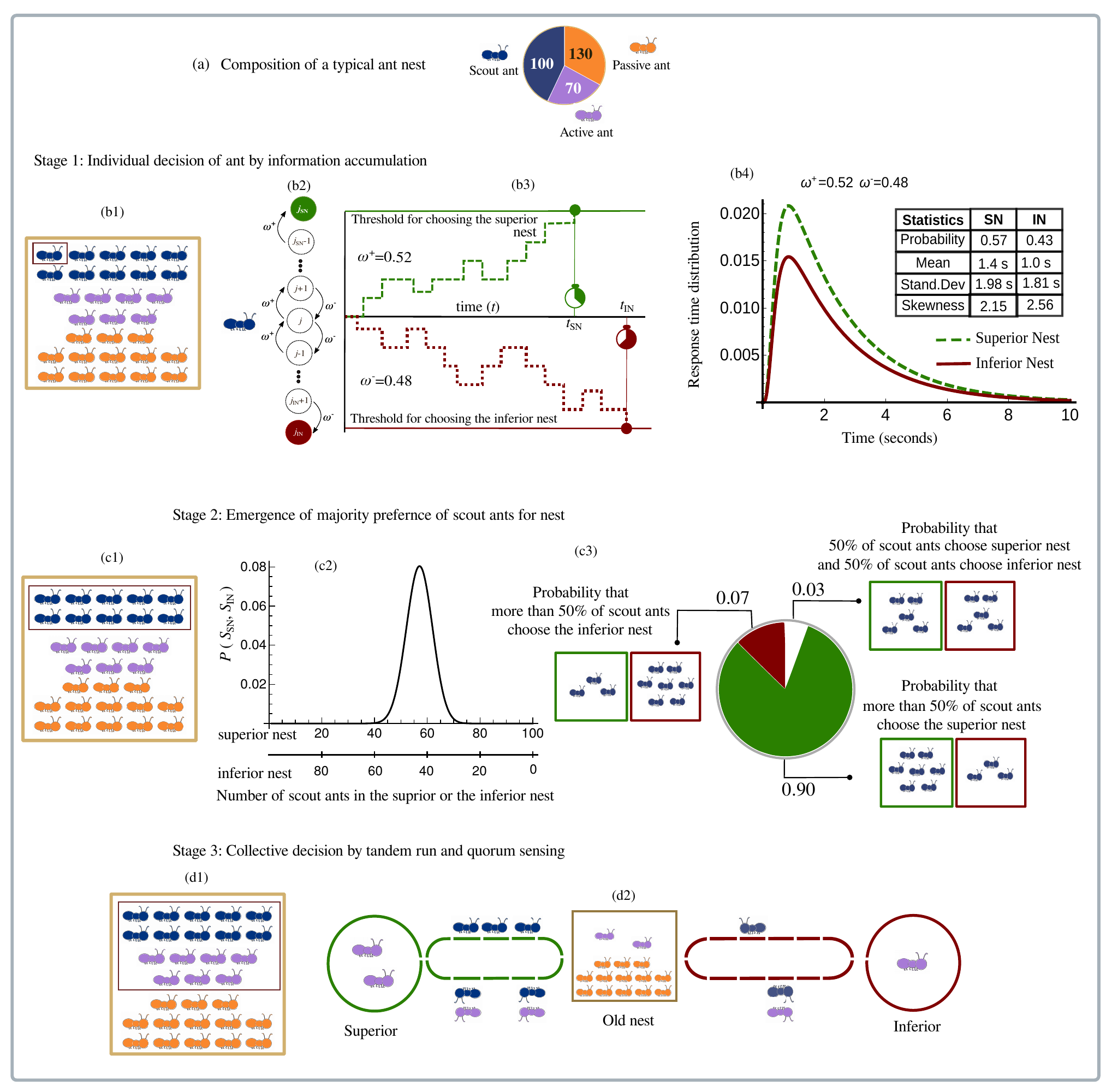}
\caption{\textbf{Multi-stage process of decision making for selecting a new nest}: (a) The ant population in the original nest comprises of three kinds of ants. Scout ants assess multiple sites before arriving at their respective individual choices for the potential new nest site.  Active ants  then follow scout ants, via tandem runs, to the sites chosen by the leading scout. After the collective decision emerges, passive ants are carried by the scout and active ants to their new nest. Although the model and analytical results are general, for graphical presentation of the typical results  we choose a specific set of values of  the nest population composition: 300 ants consists of 100 scout ants, 70 active ants and 130 passive ants; this choice  is motivated by the corresponding values observed in  experiments \cite{pratt05}. (b1)-(b3) (Stage 1) The drift-diffusion model describes the stochastic process of `information' collection by a scout ant. When the accumulated information first attains a certain pre-assigned threshold for one of these sites that site becomes its own individual choice; thus time taken by a scout to make its individual choice is a {\it first-passage time}.
 (b4) The distribution of the individual decision time for choosing the superior (green dashed line) and inferior (red solid line) nests.  The parameter values taken are $j_0=5$, $j_\text{SN}=10$, $j_\text{IN}=0$, $\omega^+=0.52$ and $\omega^-=0.48$. For the chosen set of parameter values, the probabilities of choosing the superior and inferior sites, irrespective of the time taken, are 0.57 and 0.43, respectively, as tabulated in the inset where first few moments of the distribution are also given.  (c1)(Stage 2) Schematic depiction of scout ants that have already chosen a site for which they would advertise among the active nest mates. (c2) Probability distribution for finding $S_\text{SN}$ scout ants in the superior nest and $S_\text{IN}$ scout ants in the inferior nest at the end of Stage 2. (c3) Pie chart, for the chosen model parameter values, indicating the probability of finding more than half of the scout ants in superior nest or the inferior nest, and probability of finding equal scout ants in both superior nest and inferior nest at the end of Stage 2. (d1)-(d2) (Stage 3) Schematic depiction of the recruitment of active ants by scout ants to the new site via tandem run on a periodic trail.  Upon its arrival at the new site, each active ant decides whether to move in based on the quality and population of scout ants at that site. (d3) This recruitment is continues till a threshold number of active ants (`quorum') is achieved at either of the two sites. 
}
\label{Stage_DM}
\end{figure*}

\begin{table*}
\caption{Statistics of individual decision time of a single scout ant} 
\centering 
\begin{tabular}{c | c }
\hline
\multicolumn{2}{c} {1. Individual decision time distribution for choosing the} \\
\hline
& \\
 superior nest=$p_\text{rt}^\text{SN}(t)$ & inferior nest =$p_\text{rt}^\text{IN}(t)$ \\  & \\  =$\prod^{ d_\mathbb{W}}_{(d_\mathbb{W}^\text{SN}-1)}\omega^+ \sum_{k=1}^{d_\mathbb{W}}\left[\frac{\prod_{m=1}^{d_{\mathbb{W}}^\text{SN}}(\sigma^\text{SN}_m-\lambda_k)}{\prod_{n=1,n\neq k}^{d_\mathbb{W}}(\lambda_n-\lambda_k)} e^{-\lambda_k t}\right]$ & =$\prod^{ d_\mathbb{W}}_{(d_\mathbb{W}^\text{SN}-1)}\omega^+ \sum_{k=1}^{d_\mathbb{W}}\left[\frac{\prod_{m=1}^{d_{\mathbb{W}}^\text{SN}}(\sigma^\text{SN}_m-\lambda_k)}{\prod_{n=1,n\neq k}^{d_\mathbb{W}}(\lambda_n-\lambda_k)} e^{-\lambda_k t}\right]$ \\  & \\
where $d^\text{SN}_\mathbb{W}=j_0-j_\text{IN}-1$, & where $d^\text{IN}_\mathbb{W}=j_\text{SN}-j_0$, \\  & \\
  $\lambda_k=1+2\sqrt{\omega^-\omega^+} \cos \left( \frac{k \pi}{d_{\mathbb{W}}+1}\right )$, &  $\lambda_k=1+2\sqrt{\omega^-\omega^+} \cos \left( \frac{k \pi}{d_{\mathbb{W}}+1}\right )$,\\  & \\
 and $\sigma^\text{SN}_m=1+2\sqrt{\omega^-\omega^+} \cos\left (\frac{m\pi}{d_{\mathbb{W}}^\text{SN}+1}\right )$ & and $\sigma^\text{IN}_m=1+2\sqrt{\omega^-\omega^+} \cos\left (\frac{m\pi}{d_{\mathbb{W}}^\text{IN}+1}\right )$\\  &\\
\hline

\hline
\multicolumn{2}{c} {2. $j^{th}$ moment of the individual decision time  for choosing the} \\
\hline \\
&\\
superior nest = $\langle t^j_\text{SN} \rangle $ & inferior nest = $\langle t^j_\text{IN} \rangle $\\ & \\
$=\int _0 ^\infty t^j_\text{SN} ~ p_{rt} (t_\text{SN}) ~ d t_\text{SN} =\sum_{k=1}^{d_\mathbb{W}} \frac{C_\text{SN}(k) \Gamma(j+1)}{\lambda_k^{j+1}}$ & =$\langle t^j_\text{IN} \rangle =\int _0 ^\infty t^j_\text{IN} ~ p_{rt} (t_\text{IN}) ~ d t_\text{IN} =\sum_{k=1}^{d_\mathbb{W}} \frac{C_\text{IN}(k) \Gamma(j+1)}{\lambda_k^{j+1}} $\\ 
 & \\
 where $C_\text{SN}(k)=\prod^{ d_\mathbb{W}}_{(d_\mathbb{W}^\text{SN}-1)}\omega^+ \left[\frac{\prod_{m=1}^{d_{\mathbb{W}}^\text{SN}}(\sigma^\text{SN}_m-\lambda_k)}{\prod_{n=1,n\neq k}^{d_\mathbb{W}}(\lambda_n-\lambda_k)} \right]$ & where $C_\text{IN}(k)=\prod^{ d_\mathbb{W}}_{(d_\mathbb{W}^\text{IN}-1)}{\omega^-} \left[\frac{\prod_{m=1}^{d_{\mathbb{W}}^\text{IN}}(\sigma^\text{IN}_m-\lambda_k)}{\prod_{n=1,n\neq k}^{d_\mathbb{W}}(\lambda_n-\lambda_k)} \right]$  \\
 &\\
\hline
\multicolumn{2}{c} {3. Mean individual decision time  for choosing the}\\
\hline \\
superior nest =$\langle t_\text{SN} \rangle $ & inferior nest  =$\langle t_\text{IN} \rangle $ \\
& \\
 $=\int _0 ^\infty t_\text{SN} ~ p_{rt} (t_\text{SN}) ~ d t_\text{SN} =\sum_{k=1}^{d_\mathbb{W}} \frac{C_\text{SN}(k)}{\lambda_k^2}$ & $ =\int _0 ^\infty t_\text{IN} ~ p_{rt} (t_\text{IN}) ~ d t_\text{IN} =\sum_{k=1}^{d_\mathbb{W}} \frac{C_\text{IN}(k)}{\lambda_k^2}$ \\ &\\
\hline
\multicolumn{2}{c} {4. Standard deviation associated with the individual decision time}\\
\hline \\
 $\sigma_{t_\text{SN}} =\sqrt{\sum_{k=1}^{d_\mathbb{W}} \frac{2C_\text{SN}(k)}{\lambda_k^3}-\left(\sum_{k=1}^{d_\mathbb{W}} \frac{C_\text{SN}(k)}{\lambda_k^2}\right)^2}$ &  $\sigma_{t_\text{IN}} =\sqrt{\sum_{k=1}^{d_\mathbb{W}} \frac{2C_\text{IN}(k)}{\lambda_k^3}-\left(\sum_{k=1}^{d_\mathbb{W}} \frac{C_\text{IN}(k)}{\lambda_k^2}\right)^2}$ \\
&\\
 \hline
\multicolumn{2}{c} {5. Skewness associated with the individual decision time}\\
\hline \\
&\\
 $\kappa_{t_\text{SN}} =\frac{1}{{\left(\sigma_{t_\text{SN}} \right)^3}} \left(\sum_{k=1}^{d_\mathbb{W}} \frac{6C_\text{SN}(k)}{\lambda_k^4}-3\langle t_\text{SN} \rangle \sigma_{t_\text{SN}} -{\langle t_\text{SN} \rangle }^3 \right)$ &  $\kappa_{t_\text{IN}} =\frac{1}{{\left(\sigma_{t_\text{IN}} \right)^3}} \left(\sum_{k=1}^{d_\mathbb{W}} \frac{6C_\text{IN}(k)}{\lambda_k^4}-3\langle t_\text{IN} \rangle \sigma_{t_\text{IN}} -{\langle t_\text{IN} \rangle }^3 \right)$\\
 &\\
\hline
\multicolumn{2}{c} {6. Probability of choosing }\\
\hline \\
 superior nest = $Q_\text{SN(1)}$ & inferior nest = $Q_\text{IN(1)}$ \\ &\\
 $=\int_0^\infty p_\text{rt}(t_\text{SN}) ~ dt_\text{SN}=\frac{|-\mathbb{W}_\text{SN}|}{|-\mathbb{W}|}\prod^{ d_\mathbb{W}}_{(d_\mathbb{W}^\text{SN}-1)}\omega^+$ & $\int_0^\infty p_\text{rt}(t_\text{IN}) ~ dt_\text{IN}=1-\int_0^\infty p_\text{rt}(t_\text{SN}) ~ dt_\text{SN}$ \\
 &\\
\hline 
\end{tabular}
\label{response_statistics}
\end{table*}

\section{Stage 1 : Individual decision of a scout ant by information accumulation }
\subsection{A drift-diffusion model }
In our model the decision making by an individual scout ant is assumed to be governed by a  sequential accumulation of relevant information where the scout ants are the information accumulators. Each scout ant surveys both the superior nest (denoted by SN) and the inferior nest (IN) and accumulates information about them (see Fig.\ref{Stage_DM}(b1)). The process of information accumulation is assumed to be identical to a drift-diffusion process, introduced by Ratcliff and co-workers \cite{ratcliff78, ratcliff08} for modelling decision making by humans. Since accumulated information is assumed to change in each step of the process by a discrete amount it is represented by a discrete variable $j$. Consequently, the information accumulation process is a biased random walk or equivalently a stochastic birth-death process as shown in Fig.\ref{Stage_DM}(b2). Since two alternative sites are available to each ant as potential new nests, we model the decision process by the scout ants as a birth-death process with two absorbing boundaries at states $j_\text{SN}$ and $J_\text{IN}$ shown in Fig.\ref{Stage_DM}(b2). The absorbing boundaries denote the thresholds of accumulated information where individual decision is made by an ant (see Fig.\ref{Stage_DM}(b3)).

Let the rates of information accumulation towards the superior nest and the inferior nest be denoted by $\omega^+$ and $\omega^-$, respectively. Note that, as in the case of homogeneous random walks, $\omega^+ + \omega^-=1$.  Getting absorbed at states $j_\text{SN}$ and $j_\text{IN}$ indicate choosing the superior and the inferior nest respectively. Through out this paper we  consider  
\begin{equation}
\omega^+=0.52 ~~\text{and}~~ \omega^-=0.48.
\label{omega_plus_minus}
\end{equation}
This choice of rates is motivated by our aim to show how a very minor bias of an individual scout ant towards the superior nest is adequate for fast and correct collective decision of the ant colony.

Let $P(j,t)$ be the probability that the amount of accumulated information retained by an ant at time $t$ is $j$. The set of coupled master equations governing the noisy decision making by a single ant is  given by 
\begin{equation}
\dot{\textbf{P}}(t)=\mathbb{W}_\text{tr} \textbf{P}(t) ~
\end{equation}
where $\textbf{P}(t)$ is a vector whose elements are $P(j,t)$ and the dot on $\textbf{P}(t)$ denotes derivative with respect to time. The transition matrix 
$\mathbb{W}_\text{tr}$ is  given by
{\small
\[\begin{bmatrix}
0 &\omega^- &0 &0 & & &\\
0 & -1 & \omega^- & 0 & 0 &   &  \\
0 & \omega^+ & -1 & \omega- & 0 & 0 & \\
\ddots &\ddots & \ddots & \ddots & \ddots & \ddots &\\
0& 0 & \omega^+ & -1 & \omega^- & 0 &0 \\
 & \ddots & \ddots & \ddots &  \ddots & \ddots\\
 &  &  0 & \omega^+ & -1 & \omega^- & 0\\
 &  & & 0 & \omega^+ & -1& 0\\
0&0& & & & \omega^+ &0\\
\end{bmatrix}\] }
which is of dimension $d_\mathbb{W} \times d_\mathbb{W}$ where
\begin{equation}
d_\mathbb{W}=j_\text{SN}-j_\text{IN}
\label{d_W}
\end{equation}

Expressions for the relevant statistical quantities related to the individual decision time of a single scout ant choosing the superior or  the inferior nest are tabulated in Table \ref{response_statistics}. The detailed derivations of these expressions are given in the Appendices \ref{a_toeplitz} and \ref{a_distribution}. 
We plot the individual decision time distribution for choosing the superior and the inferior nest in Fig.\ref{Stage_DM}(b4) for the given set of rates in equation (\ref{omega_plus_minus}) and estimate the moments of distribution like the associated mean, standard deviation and skewness which are summarised in the mini table in Fig.\ref{Stage_DM}(b4). 

Interestingly, for the pair of values of $\omega^+$ and $\omega^-$ (equation (\ref{omega_plus_minus})), we get 
\begin{equation}
Q_\text{SN(1)}=0.57  ~\text{and}~ Q_\text{IN(1)}=0.43~.
\end{equation} 
which is the probability of choosing a superior and a inferior nest, respectively, by a scout ant . The ratio of $Q_\text{SN(1)}$ to $Q_\text{IN(1)}$ is equal to the corresponding experimentally measured values reported in ref.\cite{pratt02}. The small difference between $Q_\text{SN(1)}$ and $Q_\text{IN(1)}$ indicates that a single scout ant is highly prone to make erroneous decision while surveying the available nests. Hence, it would be unwise for the ant colony to rely on the decision of any single ant. }

\begin{figure*}
\includegraphics[width=0.9\textwidth]{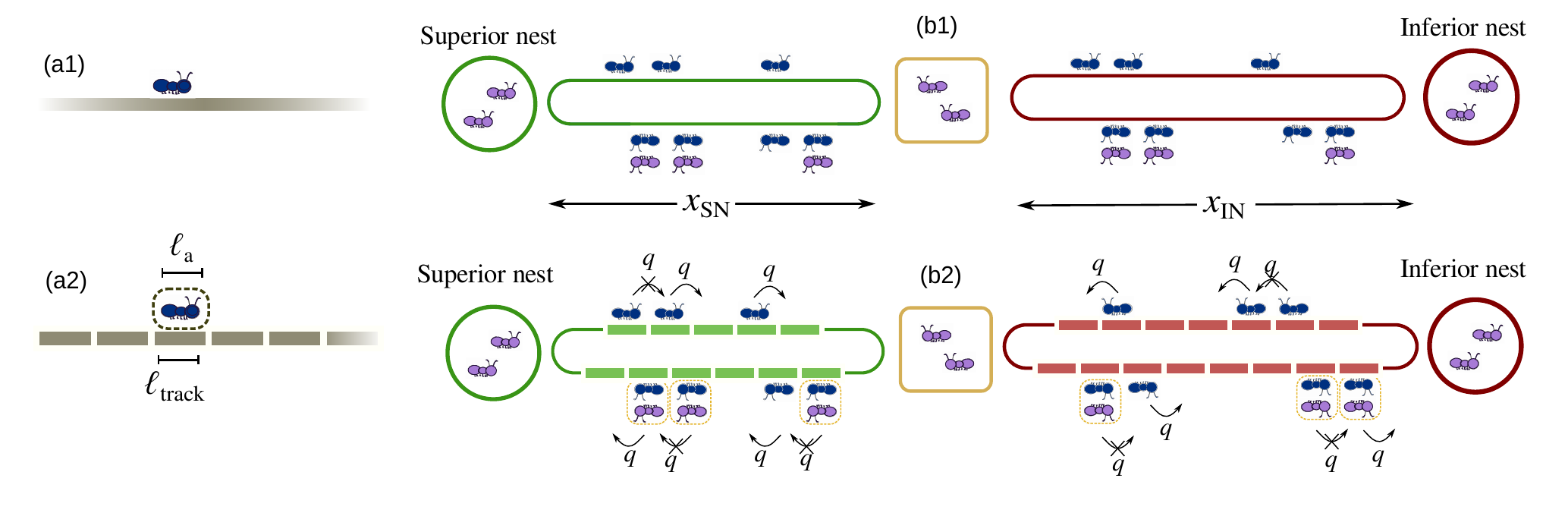}
\caption{A trail shown in (a1) is discretized and represented by a lattice chain as shown in (a2); the spacing $\ell_\text{track}$ between the successive lattice sites being equal to the linear size $\ell_\text{ant}$ (=3mm) of an ant.  For simplicity, we represent the trail joining a new nest and the old nest with a periodic tracks as shown in (b1) which are represented by lattice chains with periodic boundaries as denoted in (b2). The traffic of the ants between the old nest and a new site is modelled as a totally asymmetric simple exclusion processes (TASEP). Each individual scout ant is represented by a self driven particle. In (b2), scout ants travelling from new nest to old nest for recruiting active ants can hop to the target site with rate $q$ if the target site in front is empty. So far as the traffic from the old nest to a new site is concerned, the scout ant and the active ant following it are together modelled as a single composite particle that is assumed to hop forward at the same rate as that of a single scout ant (i.e., $q$ if the target site is empty).
  }
\label{Fig_TASEP}
\end{figure*}

\section{Stage 2 : Emergence of majority preference of scout ants for nests}
Let the total number of scout ants in the system be $S$ (see Fig.\ref{Stage_DM}(c1)). As discussed above, $Q_\text{SN(1)}$ and $Q_\text{IN(1)}$  are the probabilities of choosing the superior and the inferior nest, respectively, by an individual scout ant. Then, at the end of stage 2, the probability distribution that $S_\text{SN}$ and $S_\text{IN}$ number of scout ants have decided in favour of the superior and the inferior nests, respectively, is given by
\begin{equation}
P(S_\text{SN},S_\text{IN}) =  {S\choose S_\text{SN}} \left( Q_\text{SN(1)} \right)^{S_\text{SN}} \left( Q_\text{IN(1)} \right)^{S_\text{IN} } 
\label{binomial_dist}
\end{equation}
and $S_\text{IN}+S_\text{SN}=S$. The distribution for $S=100$ is plotted in Fig.\ref{Stage_DM}(c2).
The site which is individually selected by more than half of the total scout ant population emerges as the preference for the majority of scouts. The probability of the superior nest being prefered by majority  of the scout ants is
\begin{equation}
Q_\text{SN(S)} =\sum_{j= S/2 +1}^{S}{S\choose j}\left( Q_\text{SN(1)} \right)^j  \left( Q_\text{IN(1)} \right)^{S-j } ~.
\end{equation}
Even if the probability of selecting the superior and the inferior nest by a single scout ant is $Q_\text{SN(1)}$=0.57 and  $Q_\text{IN(1)}$=0.43, with total $S=100$ scout ants, the probability that majority of scout ants are in the superior nest is $Q_\text{SN(S)}=$0.9 (see Fig.\ref{Stage_DM} (c3)). From the binomial distribution in equation (\ref{binomial_dist}), it is estimated that the mean number of scout ants  recruiting for the superior and the inferior nest are given by $S_\text{SN}=57$ and $S_\text{IN}=43$ respectively.

\section{Stage 3: Collective decision by tandem runs and quorum sensing}
Scout ants, after deciding their individual preference  among the available sites at the end of Stage 2, next  recruit active nestmates to their respective teams (Fig.\ref{Stage_DM}(d1)). During this process, emerging from the old nest, the recruited active ant(s) follow the recruiter that travels towards the site that it has selected. This is the phenomena of tandem run as depicted in Fig.\ref{Stage_DM}(d2). The communication between the leader (recruiter) and the follower(s) is believed to be mainly through physical contact, its details are not required for our simple model. One round of the tandem run of a recruiter scout ant ends when it reaches the target site closely followed by the recruited active ant. Thus, the time needed for collective decision making becomes dependent on the traffic of scout ants on trails connecting the old nest and the two new nests. 

We model the movement of the ants on the trails as {\it total asymmetric simple exclusion process} (TASEP) \cite{Guttal}. A general track (see Fig.\ref{Fig_TASEP}(a1)) is discretized and represented by a lattice chain where the length of a lattice unit $\ell_\text{track}$ is assumed to be equal to the typical linear size of an ant $\ell_\text{ant}=3$ mm. An individual scout ant is denoted by a self driven particle (see Fig.\ref{Fig_TASEP}(a2))\cite{Wheeler1903}.  For simplicity, we represent the trail joining a new site and the old nest with a periodic track (see Fig.\ref{Fig_TASEP}(b1)). If the distance between the old nest and a new nest is $x_i$ cm ($i=$ SN or IN), the length of the periodic track is $2x$ cm. In our model, the corresponding tracks are denoted by lattice chains with periodic boundaries and contain $\approx 2x/\ell_\text{track}$ number of lattice units (see Fig.\ref{Fig_TASEP}(b2)).  

In the spirit of TASEP, no lattice site is allowed to be occupied by more than one particle simultaneously. 
All the scout ants which decided to advertise for a given nest at the end of Stage 2 are on the move and involved in the tandem runs. Hence, the number of scout ants $S_\text{SN}$ and $S_\text{IN}$ moving on the respective tracks in  periodic motion remain conserved throughout this stage. As shown in Fig.\ref{Fig_TASEP}(b2), scout ants travelling from new site to the old nest for recruiting active ants can hop to the target site with rate $q$ if it is empty. Scout ants travelling from old nest to new nest where active ants closely follow scout ants via tandem runs are assumed to be a single motile units and these units follow the same rules of volume exclusion which are followed by single scout ants. As the total number of scout ants $S_i$ moving in a given trail is fixed, the number density $\rho_i$ of the scout ants on the each of these tracks is also fixed and is given by
\begin{equation}
\rho_i=\frac{S_\text{i}}{(2x_i/\ell_\text{track})}
\end{equation}
where $i=$SN or $i=$IN. The flux of scout ants as a function of density  in each of these trails is given by:
\begin{equation}
J_i=J(\rho_i)=q\rho_i(1-\rho_i)  ~~~(i=\text{SN or IN})
\label{flux_eq}
\end{equation}
which is a fundamental result of TASEP.

\begin{figure*}
\includegraphics[width=0.7\textwidth]{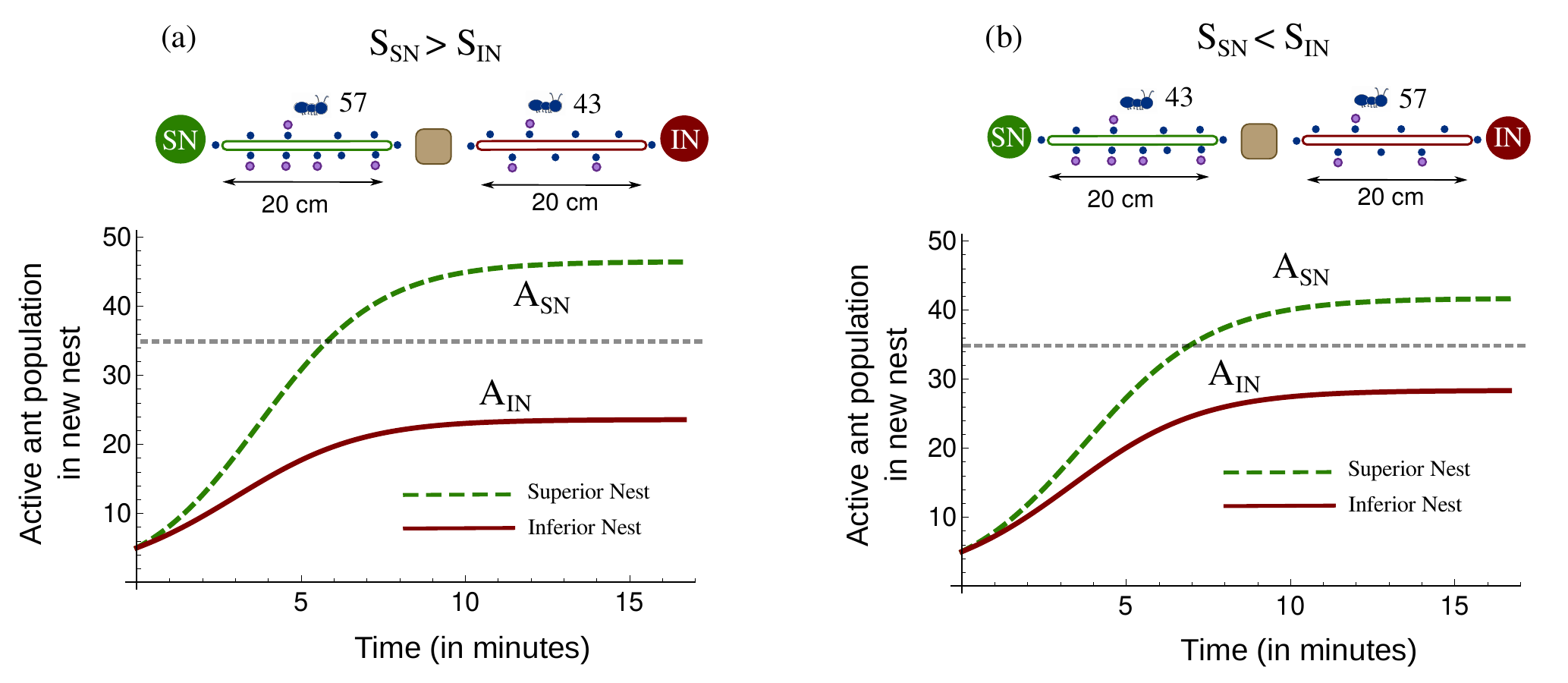}
\caption{{\bf Active ants are the ultimate decision makers:} (a) Number of scout ants recruiting for the superior nest $S_\text{SN}$=57 and the inferior nest $S_\text{IN}$=43. (b)  Number of scout ants recruiting for the superior nest $S_\text{SN}$=43 and the inferior nest $S_\text{IN}$=57. For both the cases distance between old nest and the new nests are $x_\text{SN}=x_\text{IN}$=20 cm. Number of active ants for achieving quorum at either nest is 35 as marked by the horizontal dashed line.}
\label{active_number}
\end{figure*}

\begin{figure*}
\includegraphics[width=0.7\textwidth]{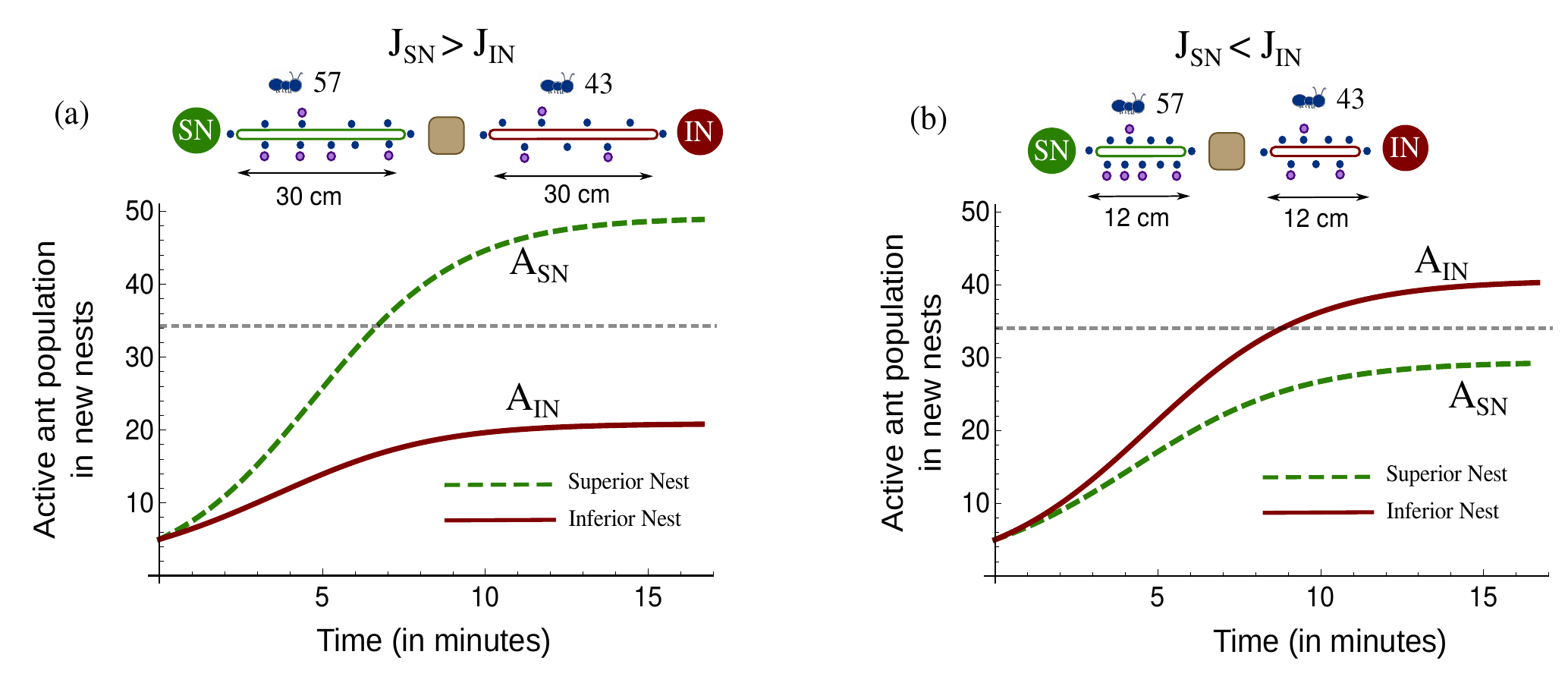}
\caption{{\bf Ant traffic during tandem runs a deciding factor:} (a) Distance between old nest and the new nests $x_\text{SN}=x_\text{IN}$=30 cm (b) Distance between old nest and the new nests $x_\text{SN}=x_\text{IN}$=12 cm. For both the cases, $S_\text{SN}$=57 and $S_\text{IN}$=43. Number of active ants for achieving quorum at either nest 35 as marked by the horizontal dashed line.}
\label{active_length}
\end{figure*}

Though the scout ants are primary assessors of the potential new site sites and recruit active ants towards their respective individual choice, the collective final decision of the colony is dependent upon the active ant population. The fluxes $J_\text{SN}$ and $J_\text{IN}$ (see equation (\ref{flux_eq})) determine the rate with which the active ants reach the new sites for inspection. After an active ant reaches a new site with a rate proportional to the flux, we assume that they accept the that site as the new site only probabilistically; the probabilities being $Q_\text{SN}(1)$ or $Q_\text{IN}(1)$, respectively, in the case of the site being superior or inferior. These probabilities are equal to those with which a scout ant makes a its individual choice between a superior and inferior site. 

We also assume that, the number of active ants which have already selected a given site as their choice for the next nest influence the newly arrived active ants with the recruiter scout  ants. Hence, the effective rate with which the population of active ants in a given potential nest site $i$ increase is proportional to the (i) flux $J_i$ of scout ants bringing the active ants by tandem runs for inspecting the site $i$, (iii) the probability of an active ant selecting the site $Q_i(1)$, (iii) the number of active ants $A_i$ who have already selected the site $i$ as their new home and (iv) also to the remaining number of active ants in the old nest $A_\text{ON}$. Those active ants who reject the  site $i$ after reaching there, return back to the old nest on their own without impacting the traffic of scout ants on the trail. Hence, the return journey is not captured by our equations. However, an active ant can make multiple trips to the same site by tandem run before accepting it as its own choice for the next nest site. 

The coupled rate equations governing the temporal evolution of the average population of active ants in the old ($A_\text{ON}$), the superior ($A_\text{SN}$) and the inferior ($A_\text{IN}$) sites are given by 
\begin{eqnarray}
\frac{dA_\text{SN}(t)}{dt} & =& J_\text{SN}~{Q_\text{SN}(1)}~ A_\text{ON}~ A_\text{SN}, \nonumber \\
\frac{dA_\text{IN}(t)}{dt} & =& J_\text{IN}~{Q_\text{IN}(1)}~ A_\text{ON}~ A_\text{IN}, \nonumber\\
\frac{dA_\text{ON}(t)}{dt} &= &-J_\text{SN}~{Q_\text{SN}(1)}~A_\text{ON}~ A_\text{SN} \nonumber\\ &&- J_\text{IN}~{Q_\text{IN}(1)}~A_\text{ON}~ A_\text{IN}
\label{Rate_active}
\end{eqnarray}
with the constraint 
\begin{equation}
A_\text{ON}(t)+A_\text{SN}(t)+A_\text{IN}(t)=A
\end{equation}
where  $A$ is the total number of active ants initially in the old nest. Solving these equations for realistic set of parameters we have two important observations:\\
(i) Active ants are the ultimate decision makers: When the number of scout ants advertising for the superior site is higher i.e, $S_\text{SN}>S_\text{IN}$ and so is the flux on the respective trails i.e, $J_\text{SN}>J_\text{IN}$, it is intutitive that the average number of active ants in the superior site will increase with a higher rate and ultimately the superior site will emerge as a winner with higher number of active ants settling for it i.e $A_\text{SN}>A_\text{IN}$ (see Fig.\ref{active_number}(a)). As estimated and summarised in Fig.\ref{Stage_DM}(c2) there is a low but still a finite probability for majority of the scout ants to prefer and advertise for the inferior site at the end of stage 2. In such case, even if the number $S_\text{SN}<S_\text{IN}$ and the flux $J_\text{SN}<J_\text{IN}$, collective decision can emerge in the favor of superior site with higher number of active ants settling for the superior site  i.e, $A_\text{SN}>A_\text{IN}$. Here the individual preference of an active ants and their influence on their peers act as the overriding factors which rules the ultimate decision  in the favor of superior site (see Fig. \ref{active_number}(b)).\\
(ii) Ant traffic during tandem runs a deciding factor: The traffic of scout ants on the trail between the old  and the new sites also plays an dominant role in the collective final decision. In Fig. \ref{active_length}(a), the higher flux of scout ants in superior site enable the active ants to arrive at the superior site more frequently for inspection. More frequent arrival and the bias towards the superior site leads towards the increment of population of active ants in the superior site which later emerges as the selected site for colony migration. On the other hand, in Fig.\ref{active_length}(b), even if the number of scout ants advertising for the superior site is higher i.e, $S_\text{SN}>S_\text{IN}$, the track lengths are selected in such a manner that the resulting flux in the corresponding tracks are $J_\text{SN}<J_\text{IN}$. Higher flux for the inferior site implies more frequent visits of the active ants towards the site and due to the minor bias between the superior and the inferior site, it is possible that the maximum number active ants ultimately settle for the inferior site and select it as their future home because it is comparatively more accessible to the active ants by tandem runs.

\section{Conclusion}

In this paper we have developed a theoretical model of the multi-stage process through which an ant colony makes a collective decision in a de-centralized manner. Specifically, we have considered the selection of one out of the two potential sites as the new nest of the ant colony. One key stage of this decision-making process involves {\it tandem run} that is known to be used by several ant species within the subfamilies Myrmicinae, Formicinae, and Ponerinae \cite{franklin14}. Interestingly, ant population in a typical nest of these species of ants is relatively small and these use tandem runs, instead of pheromone-based mechanisms, for recuitment of nest mates. Another key step in this multi-stage decision making is `quorum' which is also captured by our model. The traffic of ants between the old nest site and the potential new sites has been described as a TASEP in our model.

Using the techniques of level crossing statistics, we have calculated the statistical quantities that characterize the speed and accuracy of the decision making process of a single scout ant (see Fig.\ref{Stage_DM}(b)). Then we have demonstrated how a minor bias in favour of one of the two competing potential sites gets amplified during the different stages of decision making and ultimately leads to the emergence of the correct decision even when the individual scout ants and individual active ants are highly prone to make wrong decisions. Even if the inferior site is selected by a majority of the scout ants as their individual choice, it is still possible for the whole ant colony to collectively select the superior site through the stages of tandem runs and quorum sensing (see Fig.\ref{active_number}(a-b)). On the other hand, even if the majority of scout ants advertise for the superior site it does not guarantee that the ultimate decision will be in favour of this site because the traffic of scout ants plays a crucial role (see Fig.\ref{active_length}(a-b)).

In order to connect our theory with experimental data, we have chosen rates of information collection by the individual ants that correctly reproduce the experimentally observed probability of nest site selection by individual scout ants. Using the same set of values of the parameters we have then predicted the time needed for the emergence of collective decision of the ant colony. This estimate is very close to the corresponding experimentally measured data when the distance between the old nest and the new site is same in the theory and in the experiment.
\\

{\bf Acknowledgement}: One of the authors (DC) acknowledges support from SERB (India) through a J.C. Bose National Fellowship. 

\clearpage


\appendix

\section{Toeplitz matrix}
\label{a_toeplitz}
Toeplitz matrix is $n \times n$  triagonal matrix  given by
 \[ \begin{bmatrix}
a & b & c & 0 &   &0 \\
0 & a & b & c & 0 & \\
\ddots & \ddots & \ddots & \ddots & \ddots &\\
0 & a & b & c & 0 & \\
 & \ddots & \ddots & \ddots &  \ddots & \ddots\\
 &  &  0 & 0& a & b\\
 &  & & 0 & 0 & a\\
\end{bmatrix}\]
and its eigenvalues are given by
\begin{equation}
\lambda_k=a+2\sqrt{bc} \cos \left( \frac{k \pi}{n+1} \right) ~.
\end{equation}

\section{Detailed steps for calculating individual time distribution}
\label{a_distribution}
The transition matrix $\mathbb{W}_\text{tr}$ could be decomposed into a transient and a stationary part. The transient matrix denoted by $\mathbb{W}$ can be obtained from the transition matrix $\mathbb{W}_\text{tr}$ by deleting the first and last rows and the first and last columns. $\mathbb{W}$ is a special matrix known as the Toeplitz matrix \\
\begin{equation}
\mathbb{W}= \begin{bmatrix}
-1 & \omega^- & 0 & 0 &   &0 \\
\omega^+ & -1 & \omega^- & 0 & 0 & \\
\ddots & \ddots & \ddots & \ddots & \ddots &\\
0 & \omega^+ & -1 & \omega^- & 0 & \\
 & \ddots & \ddots & \ddots &  \ddots & \ddots\\
 &  &  0 & \omega^+ & -1 & \omega^-\\
 &  & & 0 & \omega^+ & -1\\
\end{bmatrix}
\end{equation}
which is square matrix of dimension $d_\mathbb{W} \times d_\mathbb{W}$ where 
\begin{equation}
d_\mathbb{W}=j_\text{total}-1~ . 
\label{dimW}
\end{equation}
Solving $\dot{\textbf{P}}(t)=\mathbb{W} \textbf{P}(t)$ gives us $P(j,t)$ for all $j$ ($j_\text{IN}<j<j_{SN}$).

$\mathbb{W}_\text{SN}$ is a submatrix obtained by keeping the first $d_\mathbb{W}^\text{SN}$ rows and first $d_\mathbb{W}^\text{SN}$ columns of the matrix $\mathbb{W}$.

\begin{itemize}
    \item \textbf{Step I}:\\ Write the master equation for the non-absorbing states i.e from $j_\text{IN}+1$ to $j_\text{SN}-1$ in a matrix form given by:
    \begin{equation}
    \bm{\dot p(t)}=\mathbb{W} \bm{p(t)}
    \label{mast_eq_W}
    \end{equation}
    Where $\mathbb{W}$ is a square matrix of dimension $d_\mathbb{W}=j_\text{SN}-j_\text{IN}-1$  given by:
    \[ \mathbb{W} = \begin{bmatrix}
-1 & \omega^- & 0 & 0 &   & \\
\omega^+ & -1 & \omega^- & 0 & 0 & \\
\ddots & \ddots & \ddots & \ddots & \ddots &\\
0 & \omega^+ & -1 & \omega^- & 0 & \\
 & \ddots & \ddots & \ddots &  \ddots & \ddots\\
 &  &  0 & \omega^+ & -1 & \omega^-\\
 &  & & 0 & \omega^+ & -1\\\end{bmatrix}\]
If the initial state is $j_0$, the initial condition is given by 
\begin{equation}
p_j(0)=\delta_{j,j_0}
\end{equation}
the formal solution to equation (\ref{mast_eq_W}) can be given by:
\begin{equation}
\bm{p(t)}=\exp(\mathbb{W}t)\cdot \bm{p(0)}
\end{equation}
The individual time distribution for hitting threshold $j_\text{SN}$ is given by $p_\text{rt}^\text{SN}(t)$ which can be calculated by:
\begin{equation}
p_\text{rt}^\text{SN}(t)=\omega^+p_{(j_\text{SN}-1)|j_0}
\end{equation}
Hence, to calculate the distribution $p_\text{rt}^\text{SN}(t)(t)$, we have to calculate the probability $p_{(j_\text{SN}-1)|j_0}$.  $p_{(j_\text{SN}-1)|j_0}$ is the probability that the system is in information state $j_\text{SN}-1$ given that intially it was in the state $j_0$.
\item \textbf{Step II}:
We take the Laplace transform of equation(2) to obtain:
\begin{equation}
    \hat{\bm{p}}(s)=(s\mathbb{I}-\mathbb{W})^{-1}\cdot \bm{p(0)}
\end{equation}
\item \textbf{Step III}:
To calculate $p_{j_\text{SN}-1)|j_0}$,
\begin{equation}
 \hat{{p}}_{(j_\text{SN}-1)|j_0}(s)=[(s\mathbb{I}-\mathbb{W})^{-1}]_{(j_\text{SN}-1),j_0}
\end{equation}
where $ [(s\mathbb{I}-\mathbb{W})^{-1}]_{j_\text{SN}-1),j_0}$ is the $(j_\text{SN}-1,j_0)^{th}$ element of Inverse of matrix $[s\mathbb{I}-\mathbb{W}]$.  
\item \textbf{Step IV}:
To calculate an element of the inverse matrix we have to calculate the co-factor.  Thus,
\begin{equation}
\hat{{p}}_{(j_\text{SN}-1)|j_0}(s)= \frac{1}{|(s\mathbb{I}-\mathbb{W})|}C_{j_0,(j_\text{SN}-1)}
\end{equation}
where $|(s\mathbb{I}-\mathbb{W})|$ denotes Determinant of matrix $[(s\mathbb{I}-\mathbb{W}]$, and  $C_{j_0,(j_\text{SN}-1)}$ is the $(j_0,j_\text{SN}-1)$th co-factor of matrix $(s\mathbb{I}-\mathbb{W})$. 
\item \textbf{Step V}: Calculation of co-factor\\
\textbf{Step Va}: Calculate top-left submatrix of $\mathbb{W}$ of dimension $d_{\mathbb{W}s}=j_0-j_\text{IN}-1$ which can be denoted as matrix $\mathbb{W}_\text{SN}$. Calculate eigenvalues $\sigma^\text{SN}$ of matrix $-\mathbb{W}_\text{SN}$.\\
\textbf{Step Vb}:\\
$C_{j0,(j_\text{SN}-1)}$ is given by:
\begin{equation}
C_{j0,(j_\text{SN}-1)}=\prod_{j=j_0-j_\text{IN}}^{j_\text{total}-2}\omega^+\prod_{m=1}^{d_{\mathbb{W}_\text{SN}}}(s+\sigma^\text{SN}_m)
\end{equation}
where $\sigma^\text{SN}_m$ is the $m_{th}$ eigenvalue of matrix $-\mathbb{W}_\text{SN}$.
\item \textbf{Step VI}: Calculate the eigenvalues $\lambda_\mathbb{W}$ of matrix $-\mathbb{W}$.
Putting together the co-factor we can write the probability distribution for hitting $j_\text{SN}$ as:
\begin{equation}
\hat{p_\text{rt}^\text{SN}}(s)=\prod_{j=j_0-j_\text{IN}}^{d_\mathbb{W}}\omega^+\prod_{m=1}^{d_{\mathbb{W}_\text{SN}}}(s+\sigma^\text{SN}_m)\prod_{n=1}^{d_\mathbb{W}}\frac{1}{s+\lambda_n}
\end{equation}
where $\lambda_n$ is the $n_{th}$ eigenvalue of matrix $-\mathbb{W}$.

 \item \textbf{Step VII}: Perform the inverse Laplace transform to obtain the expression for $p_\text{rt}^\text{SN}(t)$ in terms of exponential dsitribution. This can be written as:
\begin{equation}
p_\text{rt}^\text{SN}(t) =\frac{ \Lambda |{-\mathbb{W}_\text{SN}}|}{|-\mathbb{W}|}E_{d_\mathbb{W}}*R_{d_{\mathbb{W}_\text{SN}}}
\end{equation}
where 
\begin{equation}
 \Lambda=\prod_{j=j_0-j_\text{IN}}^{j-\text{total}-1}\omega^+
\end{equation}  
and 
\begin{equation}
E_l=\epsilon^{\lambda_\mathbb{W}(1)}*...*\epsilon^{\lambda_\mathbb{W}(l)}
\end{equation}
, where $\epsilon^{\lambda_\mathbb{W}(n)}(t)$ is an exponetial distribution given by 
\begin{equation}
\epsilon^{\lambda_\mathbb{W}(n)}(t)=\lambda_\mathbb{W}(n)e^{-\lambda_\mathbb{W}(n)t}
\end{equation}
Symbol * denotes convolution. Object $R_l$ is of the form:
\begin{equation}
R_l=(\delta +{\sigma^\text{SN}_1}^{-1} \delta')*.......*(\delta+{\sigma^\text{SN}_l}^{-1}\delta')
\end{equation}
where ${\sigma^\text{SN}_l}$ is the $l_{th}$ eigenvalue of matrix $\mathbb{-W}_\text{SN}$.
\item \textbf{Step VIII}: Pairing and calculating the convolution chain gives us:
\begin{equation}
p_\text{rt}^\text{SN}(t)=\prod^{ d_\mathbb{W}}_{(d_\mathbb{W}^\text{SN}-1)}\omega^+ \sum_{k=1}^{d_\mathbb{W}}\left[\frac{\prod_{m=1}^{d_{\mathbb{W}_\text{SN}}}(\sigma^\text{SN}_m-\lambda_k)}{\prod_{n=1,n\neq i}^{d_\mathbb{W}}(\lambda_n-\lambda_k)} e^{-\lambda_k t}\right]
\end{equation}
\end{itemize}


\begin{thebibliography}{}

\bibitem{evans19} N.J. Evans and E.J. Wagenmakers, {\it Evidence Accumulation Models: Current Limitations and
Future Directions, PsyArXiv} (2019)

\bibitem{wardbook} L.M. Ward, {\it Dynamic Cognitive Science} (MIT press, 2002).

\bibitem{ratcliff16} R. Ratcliff, P.L. Smith, S.D. Brown and G. McKoon. Diffusion Decision Model: Current Issues and History {\it Trends in cognitive sciences}, {\bf 20}(4), 260–281 (2016)

\bibitem{chittka09} L. Chittka, P. Skorupski and N.E. Raine, {\it Speed-accuracy tradeoff in animal decision making}, Trends in Ecol. and Evol. {\bf 24}, 400-407 (2009). 

\bibitem{heitz14} R.P. Heitz, {\it The speed-accuracy tradeoff: history, physiology, methodology, and behabour}, Frontiers in Neurosc. {\bf 8}, 150 (2014). 

\bibitem{hills15} T.T. Hills, P.M. Todd, D. Lazer, A.D. Redish, I.D. Couzin and the Cognitive Search Research Group, {\it Exploration versus Exploitation in Space, Mind and Society}, Trends in Cognitive Sci. {\bf 19}, 46-54 (2015). 

\bibitem{forstmann16} B.U. Forstmann, R. Ratcliff and E.J. Wagenmakers, {\it Sequential sampling models and cognitive neuroscience: advantages, applications and extensions}, Annu. Rev. Psychol. {\bf 67}, 641-666 (2016). 

\bibitem{clithero18} J.A. Clithero, {\it Response time in economics: looking through the lens of sequential sampling models}, J. Econ. Psychol. {\bf 69}, 61-86 (2018). 

\bibitem{addicott17} M.A. Addicott, J.M. Pearson, M.M. Sweitzer, D.L. Barack and M.L. Platt, {\it A primer on fioraging and the explore/exploit trade-ff for psychiatry research}, Neuropharmacology, {\bf 42}, 1931-1939 (2017). 

\bibitem{sasaki18} T. Sasaki and S.C. Pratt, {\it The psychology of superorganisms: collective decision making by insect societies}, Annu. Rev. Entomol. {\bf 63}, 259-275 (2018). 

\bibitem{holldoblerbookSupOrg} B. H\"olldobler and E.O. Wilson, {\it Superorganism: The Beauty, Elegance, and Strangeness of Insect Societies}, (W.W. Norton \& Co, 2009).

\bibitem{kennedybook} J. Kennedy and R.C. Eberhart, {\it Swarm intelligence}, (Academic Press, 2001).

\bibitem{bonabeaubook} E. Bonabeau, M. Dorigo and G. Theraulaz, {\it Swarm Intelligence: From Natural to Artificial Systems} (Oxford University Press, 1999).



\bibitem{mallon01} E. B. Mallon, S. C. Pratt, N. R. Franks.  Individual and collective decision-making during nest site selection by the ant Leptothorax albipennis.{\it Behav Ecol Sociobiol} {\bf 50} :352–359 (2001)
\bibitem{frank02} N.R. Franks, S. C. Pratt, E. B. Mallon, N. F. Britton ,and D. J. T. Sumpter. Information flow, opinion polling and collective intelligence in house-hunting social insects. {\it Phil. Trans. R. Soc. Lond. B} {\bf 357} 1567–1583 (2002)
\bibitem{pratt02} S. C. Pratt, E. B. Mallon, D. J. T. Sumpter, N. R. Franks. Quorum sensing, recruitment, and collective decision-making during colony emigration by the ant Leptothorax albipenn{\it Behav Ecol Sociobiol }{\bf 52}:117–127 (2002)
\bibitem{dornhaus04} A. Dornhaus, N. R. Franks, R. M. Hwakins, H. N. S. Shere, Ants move to improve: colonies of Leptothorax albipennis emigrate whenever they find a superior nest site {\it Animal Behaaviour} {\bf 67} 959e963 (2004)
\bibitem{pratt05} S.C. Pratt, D.J.T. Sumpter, E.B. Mallon and N.R. Franks 2005. An agent-based model of collective nest site choice by the ant Temnothorax albipennis.{\it Anim. Behav} {\bf 70} 1023–10
\bibitem{moglich78} M. Moglich,  Social organization of nest emigration in Leptothorax (Hym. Form.). {\it Insectes Soc.} {\bf 25}, 205–225 (1978)
\bibitem{richardson07} T.O. Richardson, P.A. Sleeman, J. M.  McNamara, A.I. Houston, N. R. Franks. Teaching with
evaluation in ants. {\it Curr. Biol.}{\bf 17} 1520–1526 (2007)
\bibitem{franklin14} E.L. Franklin. The journey of tandem running: the twists, turns and what we have learned. {\it Insect Soc} {\bf 61}:1-8 (2014)
\bibitem{robinson09} Do ants make direct comparisons? E. J. H. Robinson, F. D. Smith, K. M. E. Sullivan
and N. R. Frank. {\it Proc. R. Soc. B } {\bf 276} 2635–2641 (2009)
\bibitem{ratcliff78} R. Ratcliff A theory of memory retrieval. {\it Psychol. Rev.}{\bf 85},
59–108 (1978)
\bibitem{ratcliff08} R. Ratcliff, G. McKoon. The Diffusion Decision Model: Theory and Data
for Two-Choice Decision Tasks {\it Neural Computation} {\bf 20}: 873–922 (2008)
\bibitem{ashcroft15}  Ashcroft P, Traulsen A, Galla T. When the mean is not enough: Calculating fixation time distributions in birth-death processes. {\it Phys Rev E Stat Nonlin Soft Matter Phys.} {\bf 92}(4):042154 (2015)
\bibitem{noschese06} T.S. Noschese, L. Pasquini and L. Reichel .Tridiagonal Toeplitz Matrices: Properties and Novel Applications {\it Numer. Linear Algebra Appl.}  {\bf 0}:0–0 (2006)

\bibitem{Guttal} D. Chowdhury, V. Guttal, K. Nishinari, A. Schadschneider. A cellular-automata model of flow in ant trails: Non-monotonic variation of speed with density. {\it J. Phys. A. Math. Gen.}  {\bf 35}, L573 (2002)

\bibitem{guillet19} Guillet A, Roldan E,  Julicher F, Extreme-Value Statistics of Stochastic Transport Processes: Applications to Molecular Motors and Sports   arXiv:1908.03499v3 (2019)
 
\bibitem{choi03} Choi C, Nam D Some boundary-crossing results for linear diffusion processes. {\it Statistics \& Probability Letters} {\bf 62}(3) (2003)

\bibitem{salminen88} Salminen P. On the First Hitting Time and the Last Exit Time for a Brownian Motion to/from a Moving Boundary {\it  Advances in Applied Probability}, {\bf 20}(2) 411-426 (1988)

\bibitem{rednerbook} Redner S, A guide to first-passage processes {\it Cambridge University Press} (2001)

\bibitem{Wheeler1903}  Wheeler W.M, A revision of the North American ants of the genus Leptothorax  {\it Mayr. Proc. Acad. Nat. Sci. Phila.} {\bf 55} (1903)
 
\end{thebibliography}
\end{document}